\documentclass[10pt,a4paper,twocolumn,english,aps]{revtex4-1}
\usepackage[T1]{fontenc}
\usepackage[latin9]{inputenc}
\usepackage{xcolor}
\usepackage{amsmath}
\usepackage{amssymb}
\usepackage{graphicx}
\PassOptionsToPackage{normalem}{ulem}
\usepackage{ulem}

\makeatletter

\providecolor{lyxadded}{rgb}{0,0,1}
\providecolor{lyxdeleted}{rgb}{1,0,0}

\DeclareRobustCommand{\lyxsout}[1]{\ifx\\#1\else\sout{#1}\fi}

\makeatother

\usepackage{babel}
\begin{document}
\title{A simple and general approach for reversible condensation polymerization
with cyclization}
\author{Michael Lang$^{1}$}
\email{lang@ipfdd.de}
\author{Kiran Suresh Kumar$^{1,2}$}
\affiliation{$^{1}$Institut Theorie der Polymere, Leibniz Institut für Polymerforschung
Dresden, Hohe Straße 6, 01069 Dresden, Germany}
\affiliation{$^{2}$Institut für Theoretische Physik, Technische Universität Dresden,
Zellescher Weg 17, 01069 Dresden, Germany}

\begin{abstract}
We develop a simple recursive approach to treat reversible condensation
polymerization with cyclization. Based upon a minimum set of balance
equations, the law of mass action, Gaussian chain statistics, and
the assumption of independent reactions, we derive exact analytical
solutions for systems without cyclization, for systems containing
only smallest loops, or systems that exclusively form loops. Exact
numerical solutions are computed for the general case of a homopolymerization
of flexible precursor polymers. All solutions were tested with Monte-Carlo
simulations. A generalization for good solvent is discussed and it
is shown that this generalization agrees with preceding work in the
limit of low and high polymer volume fractions. The new aspect of
our approach is its flexibility that allows for a rather simple generalization
to more complex situations. These include different kinds of reversible
linear polymerization, non-linear polymerization, first shell substitution
effect, semiflexibility, or a low molecular weight cut-off for cyclization.
\end{abstract}
\maketitle

\section{Introduction}

Polymer materials that are held together by dynamic bonds have gained
increasing interest in the past decades. These materials can provide
a reasonable compromise between durability and reusabilty \citep{Zee2020},
whereby often self-healing properties \citep{Campanella2018} come
along with simple routes for recycling \citep{Hodge2014,Bapat2020}.
The ability to respond to external stimuli through the dynamic bonds
enables a large number of applications \citep{McBride2019} where
new functionalities can be imprinted into these materials \citep{Zhang2018}.

In a linear step growth polymerization, it is inevitable that cyclic
molecules form along with linear chains \citep{Flory1953}. These
cyclic molecules (``loops'') can have quite different static \citep{Grosberg1996,Lang2012}
or dynamic \citep{Kapnistos2008,Michieletto2016} properties as compared
to their linear counterparts, in particular when being mixed with
polymers of a different architecture \citep{Zhou2019} or weight \citep{Lang2015a,Lang2013b}
or immersed in a solvent at varying concentration \citep{Lang2012}.
Cyclic molecules can concatenate \citep{GilRamirez2015} and even
may form Olympic gels under appropriate conditions \citep{Fischer2015},
while traces of long linear chains can penetrate cyclic molecules
dramatically slowing down their relaxation \citep{Kapnistos2008}.
Virtually all of these observations have been made with permanently
linked polymers, since a static polymer architecture is clearly preferable
for precise measurements of this kind. But up to the lifetime of the
bonds, the reversible systems will exhibit a similar behavior as the
irreversible one, while system properties change to the equilibrium
properties in the limit of long times. This dynamic self-adjustment
can be used for processing or when different applications on different
time scales are desired. For instance, synthesis may require a dilute
solution, producing small rings with reversible bonds. If the bonds
live longer than the time it takes to dry the polymers, drying the
solution will turn first into a low viscous melt of small rings before
it slowly adjusts its weight distribution to the higher concentration
and turns into a highly viscous melt of long chains with some rings
inside.

Reversible condensation polymerization has been discussed since more
than 70 years as one of the possible ways to incorporate dynamic bonds
into polymer materials. Not surprisingly, several excellent reviews
are available, covering different aspects of this type of polymerization.
Theory is summarized in Ref. \citep{Kuchanov2004}, while Ref. \citep{DeGreef2009}
approaches this problem from the more general viewpoint of supramolecular
polymerization. Entropy driven ring-opening polymerization is discussed
in Ref. \citep{Strandman2010}, and many kinds of polycondensation
reactions in Kricheldorf's book \citep{Kricheldorf2014}. More recently,
Di Stefano and Ercolani put a focus on their own contributions and
the concept of equilibrium effective molarity to understand ring-chain
equilibria \citep{DiStefano2016a,DiStefano2019}. In general, reversible
polymerization is one of the simplest cases of self-assembly, and
thus, an interesting model system to test and to develop theoretical
concepts. Therefore, it is rather surprising that despite of the large
progress documented in the above reviews, the theory still remains
quite incomplete.

The theory of equilibrium polymers was born with the seminal work
of Jacobson and Stockmayer (JS) \citep{Jacobson1950}. These authors
started by assuming independence of all reactions for precursor macromonomers
with Gaussian end-to-end statistics and arrived at the following key
results that endure until today. First, the weight distribution of
the linear chains remains a most probable one, independently of the
weight fraction of cyclic species and the degree of polymerization.
Second, in the limit of high conversion $p\rightarrow1$ and high
concentrations, the weight distribution of the cyclic species decays
proportional to $i^{-3/2}$, where $i$ is the number of precursor
polymers that form a cyclic molecule. Third, there is a critical concentration
below which driving the condensation to completion yields a zero weight
fraction of linear chains, while this weight fraction remains non-zero
above the critical concentration.

Qualitatively, most of these key features have been confirmed in numerous
works \citep{Jacobson1950b,Brown1965,Suter1976,Mutter1976,Petschek1986,Moratti2005,Schmid2009}
once the chains have become sufficiently long to satisfy Gaussian
statistics, see, e.g. Figure 2 of Ref. \citep*{Flory1966}. In fact,
the observed deviations were used to improve rotational isomeric state
models and similar approaches \citep{Carmichael1964,Beevers1971,Cooper1972,Flory1976b,Suter1976,Mutter1976,Sisido1976}
and helped to develop a more concise picture of chain conformations
on short length scales and the directional requirements to form cyclic
molecules. The key observation was here that the weight fraction of
cyclics formed is a measure for the probability of contacts between
the ends of the corresponding chains. This topic is still part of
current research \citep{MadeleinePerdrillat2014}, but in order to
discuss general aspects of theory in our work, we will focus on the
ideal case where the precursor polymers are sufficiently long to obey
Gaussian statistics. Similarly, we do not explicitly consider the
effect of chain stiffness \citep{Chen2004} and discuss the effect
of excluded volume only towards the end of this work.

\begin{figure}
\includegraphics[width=1\columnwidth]{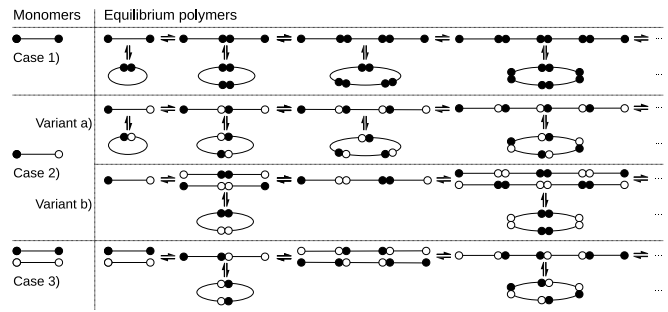}

\caption{\label{fig:The-three-classical}The three classical cases of linear
polymerization discussed in Ref. \citep{Jacobson1950} that involve
only two different reactive groups and no more than two different
precursor polymers. Reactive groups of a different type are displayed
by different beads. The different precursor molecules on the left
assemble into linear and cyclic structures where the simplest ones
are shown on the right.}
\end{figure}

The main problem of the JS approach \citep{Jacobson1950} is that
no quantitative predictions are made. For instance, conversion is
not computed, it is considered to be an adjustable parameter of theory
and must be determined post hoc from experimental data. Furthermore,
the treatment of the three classical cases sketched in Figure \ref{fig:The-three-classical}
is clearly incomplete. Here, case 1 is a homopolymerization where
all chain ends are of the same type. In case 2, each chain has two
different chain ends of type $A$ and $B$ respectively that may react
either with reactive groups of the other (variant a) or of the same
type (variant b). Case 3 is a strictly alternating copolymerization
of $A$ terminated strands with $B$ terminated strands. For case
1 and the simple variant a) of case 2 (where $A$ groups react with
$B$), the key equations were given by JS in Ref. \citep{Jacobson1950},
while variant b) of case 2 is missed and the discussion of case three
is limited to stoichiometrically balanced systems or a completely
reacted minority species. Besides these limitations, the missing quantitative
predictions for conversion or the weight fraction of rings make it
quite difficult to make accurate predictions for sample average quantities
like average molar mass, or polydispersity. This is of quite some
relevance, since it is expected that a small weight fraction of rings
at high concentrations can still cause a large polydispersity of the
sample \citep{Flory1953,Weidner2016}. We will close some of these
gaps in a subsequent work where we provide a full discussion of case
3 and variant b) of case 2 \citep{Lang2021b}.

Typical examples for case 1 are polymers that couple through their
OH groups like dimethylsilanediol \citep{Jacobson1950} or that contain
self-complementary binding sites like ureidopyrimidone dimers or guanidiniocarbonyl
pyrrole carboxylate zwitterions or alike \citep{Sijbesma1999,Hofmeier2005,Fox2009,Groeger2011,Yang2015}.
Classical examples for variant a) of case 2 include the polymerization
of some hydroxy acids like $\omega$-hydroxyundecanoic acid \citep{Flory1936,Jacobson1950}
or poly(4-hydroxybenzoate) \citep{Lieser1983}. Examples for variant
b) require two independent reactions discussed for case 1. Supramolecular
polymers of this type have attracted significant attention in recent
years \citep{Groeger2011}, since two independent mechanisms can be
addressed by an external stimulus. Finally, classical examples for
case 3 are the polymerization of some glycoles with dibasic acids
like adipic acid-decamethylene glycole \citep{Flory1936,Jacobson1950}
or various metal-ligand complexes \citep{Beck2005,Yang2015}; for
the latter, however, subsequent complexation steps are not necessarily
independent \citep{Chen2004b}. In general, any kind of telechelic
polymer with reversibly reacting groups on either end can be assigned
to one of the above cases, if the reversible groups bind exclusively
to only one other reactive group on the same chain or on a second
batch of polymers with similar properties. Thus, dynamic covalent
bonds \citep{Jin2013}, in particular those where bond breakage and
formation occurs independently, provide yet another set of examples
for which our discussion can be applied.

These problems and other difficulties for estimating the weight fraction
of rings were quickly recognized and they are still discussed controversially
in literature, see Refs. \citep{Kricheldorf2020b,Szymanski2020,Kricheldorf2020a}
for the most recent contributions. Not surprisingly, during the past
decades, several authors have tried to improve the weak points of
the JS approach.

Flory recognized \citep{Flory1953} that the conversion of the linear
species depends on the weight fraction of rings. The resulting correction
was applied once for an improved estimate, but not self-consistently
as a recursion. Ercolani et al. \citep{Ercolani1993} introduced the
missing self-consistent recursion and expressed their results as effective
molarities, which is one of the key quantities used nowadays in supramolecular
chemistry. Several reversible addition reactions were treated in Ref.
\citep{Ercolani1993} along with the simple variant of case 2 as an
example for a condensation reaction. This approach was extended in
several subsequent works by one or several of the original authors,
and it was applied, for instance, to ring-ring equilibria \citep{Ercolani2008,Cacciapaglia2009,diStefano2010},
or to polymerizations of $A-B$ strands where $A$ groups bind with
$B$ in the presence of some side reactions among the $A$ groups
\citep{DeGreef2008}. Also, random co-polymerization in the presence
of loop formation has been discussed elsewhere \citep{Szymanski1989,Szymanski1992},
and strictly alternating systems of water soluble coordination polymers
where the second ligand complex with a metal ion yields a different
binding energy were discussed with some approximations in Ref. \citep{Vermonden2003}.
But still, no general treatment of case 3 and case 2 b) could be provided.

Thermodynamics based models are a second class of approaches introduced
in literature. A Flory-Huggins lattice model was used to explore the
impact of small rings on the thermodynamics of self-assembly in Ref.
\citep{Freed2012}. An approximation for the formation of small rings
is derived and several thermodynamic properties are computed in the
presence and absence of rings. Because of the underlying lattice structure,
rings contain an even number of monomers starting from 4 units, while
these restrictions do not exist for linear chains. This causes a disassembly
of the smallest rings in the limit of low concentrations even for
a homopolymerization, which is a typical result for systems where
the shortest strands do not form rings. Nevertheless, the advantange
of an approach similar to Ref. \citep{Freed2012} is that a mean field
treatment of supramolecular systems in poor solvent becomes available.
Loverde et al. \citep{Loverde2005} derive an expression for the partition
function of a system of ideal chains where chain growth competes with
the formation of rings of arbitrary size, allowing to obtain a numerical
estimate for the weight fraction of rings. Simulation of the model
leads to a good agreement with thermodynamic quantities, but the predictive
power for the amount of rings is rather limited, and no quantitative
comparison with possible predictions for average weights of rings
or linear chains has been made.

The above discussion is complemented by approaches that have their
roots in statistical physics and the analogy between polymers and
magnetism \citep{DeGennes1972,DesCloizeaux1975}. Not long after this
analogy had been established, it was shown that equilibrium polymerization
in the absence of rings can be treated using the $n\rightarrow0$
limit of the $n$-vector model of magnetism \citep{Wheeler1980}.
Subsequently, it was identified that the $n\rightarrow1$ limit refers
to equilibrium polymers in the presence of ring formation \citep{Cordery1981,Wheeler1983},
while $n\rightarrow2$ limit describes the equilibrium of directed
polymers with ring formation \citep{Petschek1986}. These works confirm
the existence of a critical concentration \citep{Petschek1986} and
consider equilibrium polymerization as a second order phase transition
in the limit of high binding energy \citep{Semenov2014}, which is
progressively smeared out towards lower binding energies. It has been
shown that polymers with no rings, classical living polymers, and
directed living polymers form three different classes characterized
by different critical exponents and long range correlation functions
\citep{Semenov2014}. However, the scaling predictions (in particular,
regarding chain conformations) are not much different for the $n\rightarrow i$,
vector model with $i=0,1,2$ \citep{Petschek1984,Petschek1986}. This
contrasts other theoretical work \citep{Cates1990} where it was argued
that the screening of excluded volume might be strongly influenced
by polymer architecture in the vicinity of the cross-over between
the ring and the chain dominated regime.

Within the framework of the $n$-vector model, the formation of ``infinite''
molecules is considered to be impossible in the thermodynamic limit
\citep{Wheeler1992}. This argument may rule out the claim that the
thermodynamic limit would lead to the formation of short cyclic species
and one or more giant (cyclic) molecules in real systems. Nevertheless,
mean field models like Ref. \citep{Jacobson1950} may include such
a limiting behavior that is exclusively reached at $p=1$, which,
therefore, lacks any practical relevance. Interactions with a solvent
produce a rich phase behavior, allowing for tetracritical and tricritical
points, and both an upper and a lower critical solution temperature
may appear in the same mixture \citep{Corrales1998}. Semiflexibility
has also been a topic of the work related to the $n$-vector model,
in particular regarding the formation of giant micelles as one of
the best studied model systems of self-assembly \citep{Cates1990}.
Here, ring formation becomes unimportant for micelle formation at
sufficiently large persistence lengths \citep{Schoot1999}, which
has been argued also in a related simulation work \citep{Wittmer2000}.

The strength of the $n$-vector model (and related approaches) lies
in providing the correct exponents and scaling predictions. Up to
now, the weak point of these approaches remains the limitation to
case 1 and the simple variant of case 2. Also, quantitative predictions
in these latter cases are not straightforward, since typically, a
lattice based approach is chosen where loop formation within the smallest
units and sometimes even free monomers are not allowed, see for instance
Refs \citep{Cordery1981,Petschek1986,Semenov2014}. Furthermore, it
is difficult to implement polymer specific details for the smallest
rings beyond introducing a lower cut-off, reducing the applicability
of these approaches for quantitative predictions.

The above summary shows that the theoretical description is still
incomplete, and significant restrictions for a further development
of theory may depend on the particular approach chosen. From an application
oriented perspective, a versatile, simple, and extendable approach
is highly desirable. In what follows, we attempt to develop such a
theoretical platform by starting from the Jacobson Stockmayer case
of Gaussian strands that are capable to form a ring within a single
precursor macromonomer. Reactions are treated using the law of mass
action and exact solutions are presented for model cases where only
smallest loops or no loops are formed. An exact numerical solutions
for case 1 and the first variant of case 2 are discussed based upon
a minimum set of balance equations between different states of the
reacting units. We use these well known model cases to show equivalence
of our approach with preceding work. Finally, we generalize from theta
solvent to good solvent conditions and compare with Monte Carlo simulation
data regarding weight distributions and contact probabilities in good
solvent. The simulations and the numerical solutions are described
in the Supporting Information (SI). The second variant of case 2 and
case three as well as poor solvent conditions, phase separation and
fractionation will be discussed in separate works. All results are
discussed critically in the light of preceding work.

\section{\label{sec:Loop-free-case}General aspects of inter- and intra-molecular
reactions}

In what follows, we discuss only case 1 and the first variant of case
2. Since the latter is a corollary to case 1, our treatment concerns
case 1 unless stated otherwise. With the perspective of a future generalization
to reversible non-linear polymerization, let us recall that a linear
polymer is the special case of an $f$-functional star shaped polymer
with $f=2$ arms. We introduce weight fractions $\text{w}_{\text{j}}$
of these ``star'' macromonomers for a given $f$ where $j$ is the
number of reactive groups of the star that are paired with another
reactive group. For distinction, the total weight fraction of loops
is denoted by $\omega$ while weight fractions of loops made of $k$
strands is written as $\omega_{\text{k}}$.

We use $c_{\text{e}}$ to denote the concentration of the surrounding
(``external'') reactive groups attached to different molecules in
the vicinity of a given reactive group and $c_{\text{i}}$ for the
(``intra-molecular'') concentration caused by a single reactive
group of the same molecule next to a given reactive group. $c_{\text{i}}$
results from the return probability of a random walk of chain segments
that connects both reactive groups (see below). Within a mean field
approach, excluded volume of polymers, steric hindrance, correlations
between bonds, and similar effects are disregarded so that $c_{\text{e}}$
is identical to the total concentration of reactive groups in the
system.

To proceed, let us assume that all reactive groups react in pairs
and do not form larger aggregates so that $p$ is the number fraction
of the paired reactive groups among the maximum number of possible
pairs. $p$ is called the extent of reaction or ``conversion'' of
the bonds. In the absence of aggregation, the macromonomers aggregate
into molecules that are either linear chains or cyclic molecules.
Throughout this work, we call single precursor macromonomers ``strand'',
while ``molecule'' is used, if assemblies of $1,2,3,...k$ strands
are under consideration. If architecture matters, we specify this
by calling linear assemblies ``chains'' and cyclic assemblies ``rings'',
``loops'', or ``cyclic molecules''. We assume that all strands
are sufficiently long that reactions are independent of each other
(no first shell substitution effect).

First, let us consider the loop-free case as the asymptotic limit
of $c_{\text{e}}/c_{\text{i}}\rightarrow\infty$. The equilibrium
concentration of unpaired or ``open'' stickers, $c_{\text{o}}$,
is related to conversion $p$ via
\begin{equation}
1-p=\frac{c_{\text{o}}}{c_{\text{e}}}.\label{eq:1-p}
\end{equation}
The reaction constant $K$ is defined by the ratio of the concentration
of the reaction product (pairs of stickers) with a concentration $\left(c_{\text{e}}-c_{\text{o}}\right)/2$
to the concentration product of the reactants, which are the open
stickers \citep{Stukhalin2013}: 
\begin{equation}
2K=\frac{c_{\text{e}}-c_{\text{o}}}{c_{\text{o}}^{2}}=\frac{p}{\left(1-p\right)^{2}c_{\text{e}}}.\label{eq:K}
\end{equation}
The factor 2 in front of $K$ will show up below whenever the concentration
of reactive groups is twice the concentration of the corresponding
bonds (case 1 and second variant of case 2). Equation (\ref{eq:K})
can be solved for $c_{\text{o}}$, which yields 
\begin{equation}
c_{\text{o}}=\frac{\left(1+8Kc_{\text{e}}\right)^{1/2}-1}{4K}\label{eq:solution}
\end{equation}
and provides conversion through equation (\ref{eq:1-p}):
\begin{equation}
p=1-\frac{\left(1+8Kc_{\text{e}}\right)^{1/2}-1}{4Kc_{\text{e}}}.\label{eq:p-1}
\end{equation}
If only linear chains are formed, this also provides the number average
degree of polymerization,
\begin{equation}
N_{\text{n}}=\frac{1}{1-p}=\frac{4Kc_{\text{e}}}{\left(1+8Kc_{\text{e}}\right)^{1/2}-1},\label{eq:Nn1}
\end{equation}
of a linear homopolymer condensation as a function of the reaction
constant, $K$, and the concentration of the surrounding sticky groups,
$c_{\text{e}}$.

Let us introduce the weight fractions $\text{w}_{\text{j}}$ of strands
with exactly $j$ reacted (``closed'') sticky groups. In the absence
of intra-molecular reactions, $c_{\text{i}}=0$, there are $f-j$
open stickers for $\text{w}_{\text{j}}$ that allow for bond formation
proportional to the concentration of $c_{\text{e}}(1-p)$ of surrounding
stickers. Strands of weight fraction $\text{w}_{\text{j+1}}$ have
$j+1$ bonds that can break. Since two reactive groups of the same
type form a bond in case 1, we obtain $f$ balance equations that
couples the weight fraction $\text{w}_{\text{j}}$ with $\text{w}_{\text{j}+1}$
\begin{equation}
(j+1)\text{w}_{\text{j}+1}=\left(f-j\right)c_{\text{e}}\left(1-p\right)2K\text{w}_{\text{j}}\label{eq:w j+1}
\end{equation}
for all $j$ with $0\le j<f$. Note that the distribution of $\text{w}_{\text{j}}$
is normalized, and conversion can be computed from $\text{w}_{\text{j}}$
directly, see equation (3) and (4) of the SI. These latter two relations
together with the $f$ balance equations provide $f+2$ equations
for the $f+1$ weight fractions $\text{w}_{\text{j}}$ and $p$, and
thus, are exactly solvable. As solution of this set of equations,
we obtain a binomial distribution for the weight fractions 
\begin{equation}
\text{w}_{\text{j}}=\left(\begin{array}{c}
f\\
j
\end{array}\right)p^{j}\left(1-p\right)^{f-j}.\label{eq:binomial}
\end{equation}
Alternatively, this distibution can be derived using statistical arguments
if conversion $p$ is known. Finally, we have to mention that the
binomial distribution is also obtained for irreversible reaction controlled
systems in the absence of cyclization. But as soon as inter- and intramolecular
reactions compete with each other, this equivalence no longer holds.

We have used the above ideal case to test both the numerical scheme
for solving the set of balance equations and the Monte Carlo simulations
of bond breakage and formation, see section \emph{Numerical solutions}
and section \emph{Monte Carlo simulations} of the SI. Full consistency
within the numerical error of the simulations ($<10^{-3}$ for $\text{w}_{\text{j}}$)
and recursion ($<10^{-6}$ for $\text{w}_{\text{j}}$) was found indicating
a proper implementation of the numerical scheme and the simulations.

Let us now consider the opposite limit where only loops are formed
within a single strand, $c_{\text{e}}/c_{\text{i}}\rightarrow0$.
In this limit, $\text{w}_{\text{j}}\equiv0$, if $j$ is odd. Let
$z$ denote the maximum number of loops that can be placed on an $f$-functional
macromonomer, i.e. $z=f/2$ if $f$ is even and $z=(f-1)/2$, if $f$
is odd. Altogether, we have $z$ balance equations for the transition
between stars with an increasing number of loops and an even number
of closed bonds. For stars with $k$ loops, $\text{0}\le k\le z-1$,
these balance equations can be written as
\begin{equation}
\left(k+1\right)\text{w}_{2\text{k}+2}=D\left(f,j\right)c_{\text{i}}2K\text{w}_{2\text{k}}.\label{eq:wk2}
\end{equation}
Here, 
\begin{equation}
D\left(f,j\right)=\left(f-j\right)\left(f-j-1\right)/2\label{eq:degeneracy}
\end{equation}
is the degeneracy of the open sticker pairs on an $f$-functional
macromonomer with $j=2k$ bonds closed. $k+1$ on the left hand side
of equation (\ref{eq:wk2}) is the number of bonds that can break
in macromonomers with $k+1$ loops.

For simplification, we start with conditions where the Gaussian end-to-end
distribution provides reasonable estimates for $c_{\text{i}}$ (linear
Gaussian strand approximation, LGS). This is the case in a good approximatino
for stars with a large degree of polymerization in a melt or a $\theta$-solvent.
Generalization to good solvent conditions is described in the section
\emph{Good solvent.} The total number of Kuhn segments per star, $N$,
is equally distributed among the $f$ star arms, whereby the sticky
groups are located on the arm ends. Then, one obtains for a single
pair of sticky groups of the same star that
\begin{equation}
c_{\text{i}}\approx\left(\frac{3f}{4\pi Nb^{2}}\right)^{3/2}\label{eq:ci}
\end{equation}
by considering the return probability of the random walk of $2N/f$
Kuhn segments that connects both sticky groups \citep{Kuhn1934,Rubinstein2003}.

Equation (\ref{eq:wk2}) shows that loop formation within a star does
not depend on conversion, in contrast to inter-molecular reactions.
Therefore, loop formation is favored in the limit of low concentrations
as long as loop formation does not involve inter-molecular bonds.
For the present work, we are particularly interested in $f=2$ with
the analytical solution
\begin{equation}
\text{w}_{0}=\frac{1}{1+c_{\text{i}}2K}\label{eq:w0 f2}
\end{equation}
\begin{equation}
\text{w}_{2}=p=\frac{c_{\text{i}}2K}{1+c_{\text{i}}2K}.\label{eq:w2 f2}
\end{equation}

\begin{figure}
\includegraphics[angle=270,width=1\columnwidth]{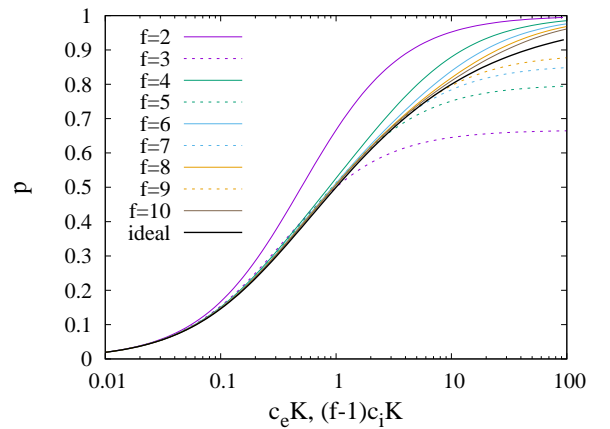}

\caption{\label{fig:Conversion-as-a-1}Conversion as a function of $(f-1)c_{\text{i}}K$
or $c_{\text{e}}K$. All lines refer to exact solutions.}
\end{figure}
The analytical solutions for $f\ge2$ are straightforward to compute,
and they are compared in Figure \ref{fig:Conversion-as-a-1} where
we have rescaled all data regarding the same effective concentration
of internal or external groups at low conversion (first loop that
is closed). These solutions differ significantly regarding conversion
from the loop-free case for small $f$, but they converge towards
it for large $f$. This observation indicates that intra-molecular
reactions may be modeled in good approximation in the limit of large
$f$ by simply subtracting the corresponding fraction of bonds from
all closed stickers. We consider this below by focussing on the small
$f$ limit when discussing systems beyond linear polymerization, $f>2$.
Note that there are also striking odd even effects at play, since
stars with even $f$ can reach full conversion, while conversion of
stars with an odd $f$ is limited by $p_{\text{max}}=1-1/f$, as the
last bond remains unpaired in the limit of $c_{\text{e}}\rightarrow0$.
For $f\ge3$, we observe an acceptable collapse of the solutions for
$p\lesssim1/2$. If the gel point conversion, $p_{\text{c}}$, remains
in this range, odd even effects will not largely contribute to a shift
of $p_{\text{c}}$ besides the point that intra-molecular reactions
are not the dominant contribution to the delay of the gel point anyway
\citep{Lang2020}.

In following sections, we focus on reaction constants in the range
of $10^{3}\le10^{7}$ and number concentrations (per unit volume $b^{3}$)
in a range from $10^{-4}<c_{\text{e}}<1$ for plotting theoretical
predictions in accord with similar plots in several reviews \citep{Brunsveld2001,DeGreef2009,Winter2016}.
$c_{\text{i}}$ is typically chosen as $10^{-2}$. This choice of
parameters allows to access the limits $c_{\text{e}}/c_{\text{i}}\gg1$
and $c_{\text{i}}/c_{e}\gg1$ and the limits of a largely smeared
out polymerization transition at low $c_{\text{e}}K$ towards a nearly
sharp transition for the largest $c_{\text{e}}K.$ Examples from literature
provide parameters in a similar range: multiple hydrogen binding units
have association constants in the range of $10^{2}$ up to $10^{9}$
$\text{M}^{-1}$, while metal coordination bonds provide association
contstants ranging from $10^{3}$ $\text{M}^{-1}$ up to $10^{21}$
$\text{M}^{-2}$ or $10^{17}$ $\text{M}^{-3}$ \citep{Yang2015}.

In general, inter- and intra-molecular reactive groups compete for
reaction with a given reactive group. Both kinds of reactions suppress
the opposite type: once a bond between two strands is formed, the
probability for ring closure is reduced due to the lower return probability
of the longer strand; similarly, intramolecular reactions in the neighborhood
reduce the concentration of surrounding reactive groups to combine
with. To understand this competition, we start with a model system,
where only the formation of the smallest loop is allowed in addition
to a linear polymerization. The advantage of this model system is
that it can be solved analytically, while it contains already the
key features of a ring-chain equilibrium.

\section{Linear polymerization with only smallest loop\label{sec:Case-1-with}}

If only the smallest loop made of exactly one strand are allowed,
there is $\omega=\omega_{1}$. At steady state, we expect the following
transitions. First, the decay rate of rings made of a single strand
must be balanced by the ring formation arising here only from strands
with zero bonds,
\begin{equation}
\omega_{1}=c_{\text{i}}2K\text{w}_{0},\label{eq:pi}
\end{equation}
see equation (\ref{eq:wk2}). The second reaction involving strands
with zero bonds is the formation of a bond with another strand and
its backwards reaction

\begin{equation}
\text{w}_{1}=2c_{\text{e}}\left(1-p\right)2K\text{w}_{0},\label{eq:w1}
\end{equation}
see equation (\ref{eq:w j+1}). Note that we assume the same reaction
constant $K$ for both inter- and intra-molecular reactions, which
is readily generalized, as these two different kinds of reactions
are described by different balance equations. Since only the smallest
loop forms in this model case, the only possible reaction for strands
with one bond closed is the formation of an inter-molecular bond.
These are balanced by the decay of inter-molecular bonds, which are
for this model system all $\text{w}_{2}$ except for the strands that
are part in loops. Similar to above, this leads to

\begin{equation}
2\left(\text{w}_{2}-\omega\right)=c_{\text{e}}\left(1-p\right)2K\text{w}_{1}.\label{eq:w2-1}
\end{equation}
Note that the modeling of a first shell substitution effect with the
above equations is straightforward, since the balance equations are
established depending on number and type of existing bonds of a given
strand. The three $\text{w}_{\text{j}}$ with $\omega$ and $p$ in
the above equations are five unknowns. The missing two equations to
solve for all unknowns are the definition of conversion and the normalization
condition of the weight fractions $\text{w}_{\text{j}}$. These can
be found in section \emph{Numerical solutions} of the SI, where we
describe briefly how a numerically exact solution of the above and
of similar sets of equations can be obtained by a simple recursion.
We compute this numerical solution along with the exact solution (see
below) in order to test the quality of the numerical solutions.

Let us proceed with the exact solution of the above smallest loop
model and discuss its properties in more detail. Within the loop fraction,
all possible bonds are closed, i.e. the conversion of loops is complete,
$p_{\text{loops}}\equiv1$, while all open stickers form the ends
of linear chains. Thus, the conversion of the linear chain fraction
\citep{Fischer2015}, 
\begin{equation}
p_{\text{lin}}=\frac{p-\omega}{1-\omega},\label{eq:p_lin}
\end{equation}
is a function of the weight fraction of loops, $\omega$, and smaller
than total conversion $p$ once $\omega>0$. For large weight fractions
of loops, $\omega\approx1$, this relation may lead to a significant
suppression of $p_{\text{lin}}$ below $p$, which counteracts the
formation of long linear chains, see equation \ref{eq:nk} below.
Therefore, it is not a priori clear whether one can expect in general
that $\omega\rightarrow1$ for $p\rightarrow1$ or whether there is
a critical concentration above which increasing $p$ will not enforce
$\omega\rightarrow1$. Indeed, such a critical behavior has first
been proposed in Ref. \citep{Jacobson1950}, and several authors have
brought good arguments for it \citep{Petschek1986,Ercolani1993,Moratti2005},
but still publications appear with an opposite point of view \citep{Kricheldorf2010,Kricheldorf2014,Kricheldorf2020,Szymanski2020}.

The reactions among reactive groups within the linear fraction of
polymers, $1-\omega$, are independent of each other. Thus, the contribution
of linear polymers to the distribution of strands with $j$ bonds,
$\text{w}_{\text{j}}$, is described by a binomial distribution, while
the loops contribute only to $\text{w}_{2}$. Therefore, we obtain
for the weight fractions $\text{w}_{\text{j}}$

\begin{equation}
\text{w}_{0}=\left(1-p_{\text{lin}}\right)^{2}\left(1-\omega\right)\label{eq:w0 bin}
\end{equation}
\begin{equation}
\text{w}_{1}=2\left(1-p_{\text{lin}}\right)p_{\text{lin}}\left(1-\omega\right)\label{eq:w1 bin}
\end{equation}
\begin{equation}
\text{w}_{2}=p_{\text{lin}}^{2}\left(1-\omega\right)+\omega\label{eq:w2 bin}
\end{equation}
for an arbitrary weight fraction of loops, $\omega$.

\begin{figure}
\includegraphics[angle=270,width=1\columnwidth]{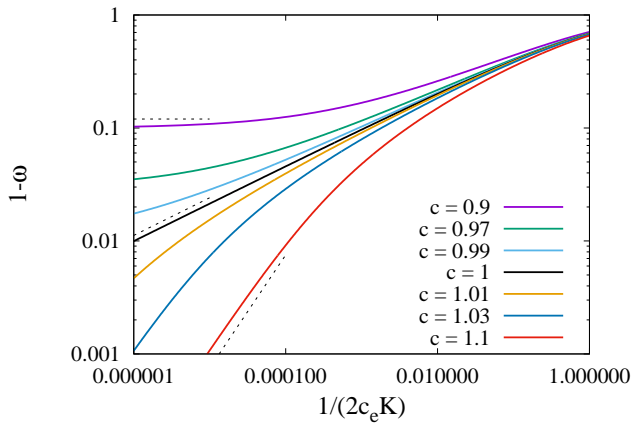}

\caption{\label{fig:Weight-fraction-of-1}Weight fraction of linear chains,
$1-\omega$, in the vicinity of the critical point for the case of
only the smallest loop. The dashed lines indicate the limiting regimes
of a linear growth, the saturation regime, and the boundary $1-\omega\propto\left(2c_{\text{e}}K\right)^{-1/3}$
for $c=1$ between the linear and the saturation regime.}
\end{figure}

Rings do not affect the distribution of the linear chains except for
consuming a portion $\omega$ of the reactive groups \citep{Jacobson1950,Wittmer2000,Fischer2015}.
This is accounted for by equation \ref{eq:p_lin}. Thus, we consider
the law of mass action for the linear species alone, as if the rings
would be inert polymers. Using equation (\ref{eq:p-1}) for case 1,
we obtain
\begin{equation}
p_{\text{lin}}=1-\frac{\left(1+8Kc_{\text{e}}\left(1-\omega\right)\right)^{1/2}-1}{4Kc_{\text{e}}\left(1-\omega\right)}.\label{eq:plin alon}
\end{equation}
Furthermore, $\omega$ is related to the conversion of the linear
species for only the smallest loop by 
\begin{equation}
\omega=c_{\text{i}}2K\text{w}_{0}=2c_{\text{i}}K\left(1-p_{\text{lin}}\right)^{2}\left(1-\omega\right),\label{eq:omega von plin}
\end{equation}
see equation (\ref{eq:pi}) and equation (\ref{eq:w0 bin}). This
quadratic equation has one solution in the interval $[0,1]$ given
by
\begin{equation}
p_{\text{lin}}=1-\left(\frac{\omega}{2c_{\text{i}}K\left(1-\omega\right)}\right)^{1/2}.\label{eq:plin2}
\end{equation}
We equate this solution with equation (\ref{eq:plin alon}) and obtain
after some algebra 
\begin{equation}
\left(1-\omega\right)\left(c-\omega\right)^{2}/\omega=\frac{c}{2c_{\text{e}}K},\label{eq:cubic}
\end{equation}
where $c=c_{\text{i}}/c_{\text{e}}$ is the concentration ratio. This
equation can be used \footnote{The analytic solution of this equation is not reproduced here because
of an unexpected degree of complexity.} to compute the weight fraction of linear chains, $1-\omega$, as
a function of $2c_{\text{e}}K$ for a given concentration ratio $c$,
see Figure \ref{fig:Weight-fraction-of-1}. Since $\omega$ is exactly
known from equation (\ref{eq:cubic}) for given concentrations $c_{\text{i}}$
and $c_{\text{e}}$ and reaction constant $K$, we can compute $p_{\text{lin}}$
and $p$ and subsequently all properties of the linear fraction and
the whole mixture of chains and rings.

Equation (\ref{eq:cubic}) shows that in the limit of $K\rightarrow\infty$,
there must be zero linear chains, $1-\omega=0$ for $c>1$, i.e. for
$c_{\text{i}}>c_{\text{e}}$, since $\omega$ is bounded by one, $\omega\le1$.
Note that $1-\omega\propto K^{-1}$ in this regime for $K\rightarrow\infty$,
and that the total concentration of cyclic polymers grows proportional
to $c_{\text{e}}$.

On the other hand, for $c<1$ in the limit of $K\rightarrow\infty$,
the term $\left(c-\omega\right)^{2}\rightarrow0$ tends to zero already
at $\omega<1$, i.e. there is a weight fraction of rings, $\omega$,
with $\omega=c_{\text{i}}/c_{\text{e}}<1$. Then, the weight fraction
of linear chains, $1-\omega$, grows $\propto\left(c_{\text{e}}-c_{\text{i}}\right)/c_{\text{e}}$
and thus, linearly as a function of $c_{\text{e}}-c_{\text{i}}$ just
beyond the critical concentration for $c_{\text{e}}>c_{\text{i}}$.
In consequence, the total amount of rings remains constant for $c_{\text{e}}>c_{\text{i}}$
while $1-\omega$ approaches a constant for $K\rightarrow\infty$.

The above scaling of the ring and chain fractions in the limit of
$K\rightarrow\infty$ agrees fully with previous work \citep{Petschek1986,Ercolani1993}
and demonstrates the existence of the critical point at $c=1$. One
interesting point is that the linear chain fraction grows as $1-\omega\propto K^{-1/3}$
in the limit of of $c_{\text{e}}K\rightarrow\infty$, where $\omega\approx1$
for $c=1$. Recall that the reaction constant is typically related
to the absolute temperature through a relation of the form 
\begin{equation}
K=V_{\text{R}}\exp\left(\frac{\epsilon}{kT}\right)\label{eq:reaction constant}
\end{equation}
where $V_{\text{R}}$ is the reaction volume and $\epsilon$ is the
energy difference between the bound and the non-bound state per pair
of stickers \citep{Stukhalin2013}. This dependence could be used
to identify the critical concentration upon a variation of temperature
similar to Figure \ref{fig:Weight-fraction-of-1} even for smeared
out transitions at a finite $K$. Figure \ref{fig:Weight-fraction-of-1}
shows equation (\ref{eq:cubic}) for a range of different $c$ in
the vicinity of the critical point. The limiting power laws for the
growth of the weight fraction of linear chains as a function $K$
are reached only for sufficiently large $c_{\text{e}}K$.

The exact solution of equation (\ref{eq:cubic}) above was used as
a test of the stability of the numerical recursion. Excellent agreement
was obtained for the parameter range of interest with $K\le10^{8}$
and $10^{-4}\le c_{\text{e}},c_{\text{i}}\le1$. Numerical instabilities
are possible for very large $c_{\text{e}}K$, depending on the precision
of the real numbers and the set of parameters chosen. We have checked
this point carefully, and present below only data from stable solutions.

\begin{figure}
\includegraphics[angle=270,width=1\columnwidth]{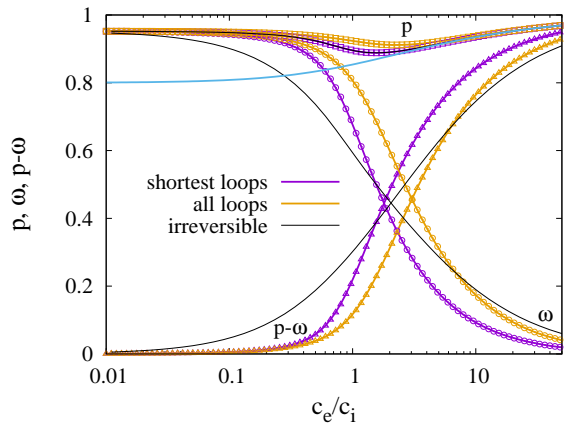}

\caption{\label{fig:-and-}Intra-molecular ($\omega$) and inter-molecular
($p-\omega$) contributions to conversion $p$ as indicated in the
Figure. Data points are Monte Carlo simulation data for $K=1000$
and $c_{\text{i}}=1/100$ for the model cases of only the smallest
loop, or of all loop sizes allowed. The lines that coincide with the
simulation data are the numerical solutions of equation (\ref{eq:pi})
to (\ref{eq:w2-1}) for the corresponding cases. The thin black lines
are obtained with equation (\ref{eq:pi-1}) and model irreversible
reactions in the case of only the smallest loop. The light blue line
is equation (\ref{eq:p-1}) where a total concentration $c_{\text{t}}=c_{\text{i}}+c_{\text{e}}$
of reactive groups is considered instead of $c_{\text{e}}$ as a simple
approximation for the limit of high concentrations.}
\end{figure}

In Figure \ref{fig:-and-}, we compare the Monte-Carlo simulation
data (see section \emph{Monte Carlo simulations} of the SI for simulation
details) with the numerical solutions for parameters that are typical
for the high conversion limit, since $p\approx0.95$ for all concentrations.
Regarding conversion, we observe a characteristic dip at the transition
from the loop dominated regime at small $c_{\text{e}}$ towards the
chain dominated regime at large $c_{\text{e}}$ that agrees with preceding
work \citep{Chen2004,Hagy2007}. Note that $\omega$ is equivalent
to the fraction of bonds among all possible bonds that are in loops
for case 1. For comparison, we have also included an estimate for
the smallest loop model in case of irreversible reactions. This estimate
is obtained by integrating the ring closure probability 
\begin{equation}
R_{\text{i}}(p)=\frac{c_{\text{i}}}{c_{\text{i}}+c_{\text{e}}(1-p)}\label{eq:pi-1}
\end{equation}
as described in Ref. \citep{Lang2005a} up to exactly the conversion
reached in the corresponding reversible case. Roughly for $c_{\text{e}}>c_{\text{i}}$,
the corresponding irreversible reaction produces more smallest loops
than in the reversible case, which is in line with the so called self-dilution
effect \citep{Kricheldorf2008}. In the low concentration limit, this
trend is reversed: here, loop formation dilutes the chains and enforces
further loop formation. Altogether, loop formation enhances further
loop formation while chain growth enhances further chain growth.

\section{Homopolymerization with cyclization (case 1) \label{sec:Case-1-with-1}}

The balance equations of the preceding section were written in a form
that only a relation between $\omega$ and $\omega_{1}$ is required
to generalize to arbitrary loop sizes. We compute $\omega$ by considering
the number fraction of loops made of $k$ strands, $l_{\text{k}}$.
These number fractions couple to the number fraction of $k$-chains,
\begin{equation}
n_{\text{k}}=\left(1-p_{\text{lin}}\right)p_{\text{lin}}^{k-1},\label{eq:nk}
\end{equation}
through the balance equations
\begin{equation}
kl_{\text{k}}=c_{\text{i}}2Kk^{-3/2}n_{\text{k}}.\label{eq:balances}
\end{equation}
The factor $k^{-3/2}$ arises from the return probability of a random
walk made of $k$ linear strands, while the factor $k$ on the left
is due to the $k$ bonds in the ring that can break at any instance
of time. Note that equation (\ref{eq:balances}) is the starting point
for generalizations where the formation of shortest loops for a specific
polymer, an explicit chain stiffness, or effects of excluded volume
or interactions with a surrounding solvent are taken into account.
Equation (\ref{eq:balances}) is rearranged to provide the JS size
distribution of loops 
\begin{equation}
l_{\text{k}}=c_{\text{i}}2Kk^{-5/2}\left(1-p_{\text{lin}}\right)p_{\text{lin}}^{k-1},\label{eq:lk}
\end{equation}
with the characteristic $k^{-5/2}$ power law times an exponential
cut-off arising from the most probable weight distribution of the
linear chains. Since $k=1$ refers to equation (\ref{eq:pi}), we
obtain
\begin{equation}
\omega=\omega_{1}\sum_{k=1}^{\infty}k^{-3/2}p_{\text{lin}}^{k-1}.\label{eq:wO}
\end{equation}
In comparison with equation (\ref{eq:balances}), we see that the
weight distribution of rings provides a direct measure for the return
probability of the random walks as a function of $k$. This point
has been used previously to test models on chain conformations as
we have mentioned in the introduction. Note that $p_{\text{lin}}<1$,
therefore, all the larger loops $k>1$ contribute only a numerical
coefficient of order unity to $\omega_{1}$, which does not surpass
$\sum_{k=1}^{\infty}k^{-3/2}\approx13/5$. Thus, we expect no modifications
of the scaling discussion of the preceding section except for a small
shift of the critical point by a factor $\omega/\omega_{1}$ to a
critical concentration of 
\begin{equation}
c_{\text{e}}=c_{\text{crit}}\equiv\lim_{p_{\text{lin}\rightarrow1}}\left(c_{\text{i}}\omega/\omega_{1}\right)\approx\frac{13}{5}c_{\text{i}}\label{eq:c_all}
\end{equation}
as proposed already in preceding work \citep{Jacobson1950b,Petschek1986,Ercolani1993}.
Altogether, equation (\ref{eq:wO}) is the missing relation between
$\omega$ and $\omega_{1}$ and the only difference between the shortest
loop model and an unrestricted model with loops of arbitrary sizes.
The above equations can be solved using our numerical scheme described
in section \emph{Numerical solutions} of the SI, providing $\omega$
and $p_{\text{lin}}$ from which all distributions can be computed.
Still, the distributions need to be normalized to give weight fractions
of $\omega$ and $1-\omega$ respectively.

All averages discussed below are available from these distributions
or can be computed for known $p$ and $\omega$ using the following
equations. The average degree of polymerization of the linear chains
is

\begin{equation}
N_{\text{n,lin}}=\frac{1}{1-p_{\text{lin}}}=\frac{1-\omega}{1-p},\label{eq:Nnlin-1}
\end{equation}
and the weight average degree of polymerization is
\begin{equation}
N_{\text{w,lin}}=\frac{1+p_{\text{lin}}}{1-p_{\text{lin}}}=\frac{1+p-2\omega}{1-p},\label{eq:Nwlin}
\end{equation}
leading to a polydispersity of the linear chains of
\begin{equation}
\frac{N_{\text{w}}}{N_{\text{n}}}=1+p_{\text{lin}}=\frac{1+p-2\omega}{1-\omega}.\label{eq:polydis}
\end{equation}
For the loops, we obtain the corresponding quantities numerically
through summation over the loop size distribution
\begin{equation}
N_{\text{n,loops}}=\frac{\sum_{k=1}^{\infty}kl_{\text{k}}}{\sum_{k=1}^{\infty}l_{\text{k}}}\label{eq:Nnloops-2}
\end{equation}

\begin{equation}
N_{\text{w,loops}}=\frac{\sum_{k=1}^{\infty}k^{2}l_{\text{k}}}{\sum_{k=1}^{\infty}kl_{\text{k}}}.\label{eq:Nwloops}
\end{equation}

For computing sample average degrees of polymerization or polydispersity,
we need to know the number density of loops, 
\begin{equation}
n_{\text{L}}=\frac{\omega}{N_{\text{n,loops}}},\label{eq:number density-1}
\end{equation}
and the number density of chains, 
\begin{equation}
n_{\text{C}}=\frac{1-\omega}{N_{\text{n,lin}}}=1-p\label{eq:NC}
\end{equation}
per original strand. The sample average degree of polymerization is
then given by
\begin{equation}
N_{\text{n}}=\frac{1}{n_{\text{L}}+n_{\text{C}}}=\frac{N_{\text{n,loops}}}{\omega+\left(1-p\right)N_{\text{n,loops}}},\label{eq:savNn}
\end{equation}
while the weight average degree of polymerization is computed through
\begin{equation}
N_{\text{w}}=\omega N_{\text{w,loops}}+\left(1-\omega\right)N_{\text{w,lin}}\label{eq:SAVNw}
\end{equation}
\[
=\omega N_{\text{w,loops}}+\frac{\left(1-\omega\right)\left(1+p-2\omega\right)}{1-p}.
\]

\begin{figure}
\includegraphics[angle=270,width=1\columnwidth]{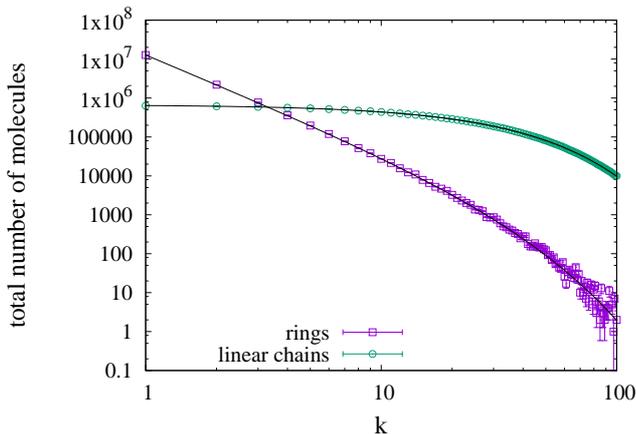}

\caption{\label{fig:Size-distributions-of}Number distributions of linear chains,
$n_{\text{k}}$, and rings, $l_{\text{k}}$, for $c_{\text{e}}/c_{\text{i}}=30$,
and $c_{\text{i}}=10^{-2}$ and $K=10^{3}$. The data points are simulation
data, while the lines are exact numerical solutions multiplied by
the total number of molecules. The numerical solution provides here
$p=0.961$, $p_{\text{lin}}=0.959$, and $\omega=0.0649$.}
\end{figure}

An example for the numerical solution with all loop sizes allowed
is included in Figure \ref{fig:-and-}. According to equation (\ref{eq:c_all}),
the transition from the loop to the chain dominated regime is shifted
to higher concentrations. The dip in conversion at $c_{\text{e}}/c_{\text{i}}\approx1$
is somewhat smeared out by the larger loops as compared to the smallest
loop model.

The number distributions of rings, $l_{\text{k}}$, and linear species,
$n_{\text{k}}$, obtained from Monte Carlo simulations follow well
the exact numerical prediction as demonstrated in Figure \ref{fig:Size-distributions-of},
where the lines are the corresponding distributions, equation (\ref{eq:nk})
and equation (\ref{eq:wO}), with $p_{\text{lin}}$ and $\omega$
as computed through the numerical solution of the equations (no fit
to data). The data in Figure \ref{fig:Size-distributions-of} is presented
in absolute numbers to provide an impression of the accuracy of the
Monte Carlo simulations.

In the original work by Jacobson and Stockmayer \citep{Jacobson1950},
it was recommended to ``guess'' $p_{\text{lin}}$ for the computation
of $\omega$, while it was also mentioned that $p_{\text{lin}}$ may
not be too far from $p$. Indeed, such a trend may be confirmed by
simulation data (see caption of Figure \ref{fig:Size-distributions-of}),
and one may put $p\approx p_{\text{lin}}$ for simplification. In
Figure 1 of the SI, we compare $p$ with $p_{\text{lin}}$ to test
whether such an approach might be useful. Indeed, $p_{\text{lin}}$
is approaching $p$ but only for $c_{\text{e}}/c_{\text{i}}\gtrsim10$.
For $2\lesssim c_{\text{e}}/c_{\text{i}}\lesssim10$, there are already
striking differences between $p$ and $p_{\text{lin}}$, while for
$c_{\text{e}}/c_{\text{i}}<2$, a largely different behavior is obtained
due to the dominant contribution of the rings to conversion.

\begin{figure}
\includegraphics[angle=270,width=1\columnwidth]{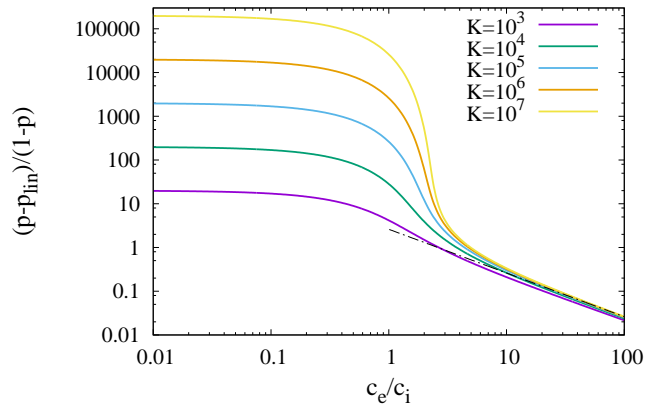}

\caption{\label{fig:The-relative-error}The relative error for estimating the
average degree of polymerization of the linear chains by assuming
$p=p_{\text{lin}}$ for $c_{\text{i}}=10^{-2}$. The dashed line is
$c_{\text{crit}}/c_{\text{e}}$, see equation (\ref{eq:plin2-1}).}
\end{figure}

The relative error for computing the average degree of polymerization
of linear chains by assuming $p_{\text{lin}}=p$ is given by the ratio
of the corresponding average degrees of polymerization minus one,
shown in Figure \ref{fig:The-relative-error}. Mathematically, we
obtain for this quantity
\begin{equation}
\frac{1-p_{\text{lin}}}{1-p}-1=\frac{\omega}{1-\omega}.\label{eq:relative error}
\end{equation}
Thus, the relative error is exactly the ratio between the fraction
of rings and the weight fraction of chains. The above relation is
rather useful, since it allows to determine the reaction constant,
see below, and it can be transformed to provide a rather accurate
estimate for $p_{\text{lin}}$ in the limit of high concentrations.

In the loop dominated case, $c_{\text{i}}\gg c_{\text{e}}$, we assume
that chains can be neglected so that we can use equation (\ref{eq:w2 f2})
with $p\approx\omega$ and equation (\ref{eq:w0 f2}) with $\text{w}_{0}\approx1-\omega$
. This provides 
\begin{equation}
\frac{\omega}{1-\omega}\approx c_{\text{i}}2K.\label{eq:low concentration}
\end{equation}
Thus, a determination of essentially the weight fraction of linear
chains, $1-\omega$, in the low concentration limit provides $c_{\text{i}}2K$
and therefore, the reaction constant, if $c_{\text{i}}$ is known
or vice versa.

In the chain dominated limit, $c_{\text{e}}\gg c_{\text{i}}$ where
$\omega\ll1$, we recall from the preceding section that the amount
of linear chains grows approximately linear with $c_{\text{e}}$,
while the amount of rings is essentially constant and almost unity
at the critical concentration. This provides $\omega\approx c_{\text{crit}}/c_{\text{e}}$
and thus,
\begin{equation}
\frac{\omega}{1-\omega}\approx\frac{c_{\text{crit}}}{c_{\text{e}}}.\label{eq:high concentration}
\end{equation}
Therefore, we can make a rather precise ``guess'' for the difference
between $p$ and $p_{\text{lin}}$ in the limit of $c_{\text{e}}\gg c_{\text{i}}$
and for large $K$: 
\begin{equation}
p_{\text{lin}}=p-\left(1-p\right)c_{\text{crit}}/c_{\text{e}},\label{eq:plin2-1}
\end{equation}
since $p$ can be estimated from equation (\ref{eq:p-1}) quite accurately
by using an effective total concentration of $c_{\text{t}}=c_{\text{e}}+c_{\text{i}}$
instead of $c_{\text{e}}$ as demonstrated in Figure \ref{fig:-and-}.
Figure \ref{fig:The-relative-error} shows, that both approximations
for equation (\ref{eq:relative error}) given above work well indeed:
equation (\ref{eq:low concentration}) is approached for sufficiently
small $c_{\text{e}}/c_{\text{i}}\lesssim10^{-1}$ while the data at
$c_{\text{e}}/c_{\text{i}}\gtrsim10^{-1}$ approaches equation (\ref{eq:high concentration})
independent of $K$.

In this latter limit $c_{\text{e}}\gg c_{\text{crit}}$ and $c_{\text{e}}K\gg1$,
one can use equation (\ref{eq:p-1}) to obtain approximations for
the number average 
\begin{equation}
N_{\text{n,lin}}\approx\left(c_{\text{e}}-c_{\text{crit}}\right)\left(\frac{2K}{c_{\text{e}}}\right)^{1/2}\label{eq:Nnlin-2}
\end{equation}
and weight average degree of polymerization
\begin{equation}
N_{\text{w,lin}}\approx2\left(c_{\text{e}}-c_{\text{crit}}\right)\left(\frac{2K}{c_{\text{e}}}\right)^{1/2}-1\label{eq:Nwlin-1}
\end{equation}
of the linear chains. Therefore, a series of (linear chain) data at
different $c_{\text{e}}\gg c_{\text{crit}}$ allows to determine both
$K$ and $c_{\text{crit}}$ simultaneously, if experimental data is
available for $N_{\text{n,lin}}$ or $N_{\text{w,lin}}$, or if experimental
data is largely dominated by the molecular weight averages of the
linear chains.

\begin{figure}
\includegraphics[angle=270,width=1\columnwidth]{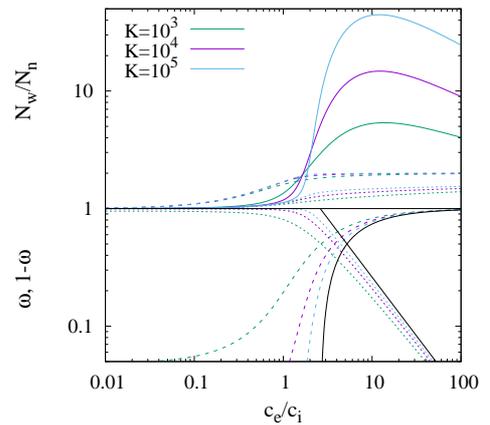}

\caption{\label{fig:Polydispersity-and-weight}Upper part: polydispersity of
full sample (continuous lines), linear chains (dashed lines), and
rings (dotted lines), for a range of reaction constants $K$ as a
function of $c_{\text{e}}/c_{\text{i}}$ with $c_{\text{i}}=10^{-2}$.
Lower part: weight fractions of rings (continuous lines) and linear
chains (dashed lines) for the same examples as in the upper half of
the Figure. The black dotted line indicates the limit of $K\rightarrow\infty$.}
\end{figure}

We conclude the discussion of case 1 by presenting numerical data
for the polydispersity of the full sample, the linear chains, and
the rings in comparison with the weight fractions of linear chains
and rings for a series of reaction constants $K$ in Figure \ref{fig:Polydispersity-and-weight}.
Typically, a distinct maximum in the polydispersity is found at $c_{\text{e}}/c_{\text{i}}\approx10$
that grows as a function of $K$. The position of the maximum is clearly
beyond the critical concentration and indicates that a significant
number fraction of linear chains is required to develop largely polydisperse
samples. This supports preceding proposals, where at high concentration,
the weight fraction of rings is expected to be small, however, their
number fraction could still be significant and have a marked impact
on the polydispersity of the sample \citep{Flory1953,Weidner2016}.
The polydispersity of the linear chains approaches two in the limit
of large concentrations and large $K$, while the polydispersity of
the rings is significantly narrower for the parameters of Figure \ref{fig:Polydispersity-and-weight}.

\section{Variant A of Case 2 \label{subsec:Case-2-R}}

Within our mean field model, the linear polymerization of monomers
$A-B$ with two different reactive groups $A$ and $B$ on either
end that form exclusively bonds between $A$ and $B$ groups (variant
1) is equivalent to case 1, except for two points. First, the external
concentration of possible reaction partners is only $1/2$ of the
total concentration of external groups, $c_{\text{e}}$. Second, any
factor of $2K$ in the balance equations reduces to $K$, since now,
the concentrations of $A$ groups, $B$ groups, and the concentration
of the bonds at full conversion are identical. Therefore, all equations
of section \emph{Homopolymerization with cyclization (case 1) }apply
after replacing $c_{\text{e}}$ by $c_{\text{e}}/2$ and $2K$ by
$K$ everywhere. In effect, the critical concentration (the polymerization
transition) requires a total concentration of reactive groups that
is larger by a factor of two as compared to case 1. Also, loop formation
is enhanced roughly by a factor of two as compared to case 1 in the
limit of high concentrations. In the low concentration limit, the
weight fraction of loops is almost identical to case 1. Note that
additional differences for such ``directional'' polymers as compared
to the ``non-directional'' polymers of case 1 have been discussed
in literature. These cover a broad range of effects that range from
minute shifts of the exponents governing conformations \citep{Petschek1984}
to completely new physics like polymer induced interactions on large
scales that do not appear for case 1 polymerization \citep{Semenov2014}.
To the best of our knowledge, the latter have not been confirmed yet
by experimental data or simulations, while the former are not substantial.
Therefore, we expect that our simple approach still provides a reasonable
approximation for variant A of case 2.

\section{Good solvent}

Similar to our discussion on polymerization, we start with considering
the precursor macromonomers as star polymers with $f=2$ arms and
we write down basic equations as a function of $f$. Furthermore,
we focus on the limit of small $f$, since the competition between
intra- and inter-molecular bond vanishes for large $f$, see section
\emph{General aspects of inter- and intra-molecular reactions}. Below,
we use the scaling relations for irreversibly bound chains for our
derivations, since it has been shown previously that conformations
of reversible chains agree (within numerical error) with the conformations
of irreversible chains \citep{Wittmer2000}. Some of the results discussed
in the following were first derived from the Potts model \citep{Wittmer1998,Wittmer2000}
or renormalization groups approaches \citep{Friedman1988,Friedman1989}.
We present a somewhat simpler derivation based upon the blob model
and contact exponents.

We consider the reactions of macromonomers in a broad range of polymer
volume fractions $\phi$ above and below their overlap polymer volume
fraction \citep{DeGennes1979}, 
\begin{equation}
\phi^{*}\equiv\frac{b^{3}N}{R_{\text{0}}^{3}}\approx\left(\frac{f}{2}\right)^{3\nu}N_{0}^{1-3\nu},\label{eq:c*}
\end{equation}
that we define here with respect to the averge distance between two
reactive groups inside the ``star'', $R_{\text{0}}$, in the limit
of zero polymer volume fraction. This distance provides a measure
for the volume from which surrounding reactive groups are chosen for
reaction. Recall that $N/f$ is the degree of polymerization of a
star arm, and $2N/f$ is the degree of polymerization of two star
arms. The good and the a-thermal regime have an upper bound that is
given by the cross-over to the concentrated regime at a polymer volume
fraction of 
\begin{equation}
\phi^{**}\approx\frac{\text{v}}{b^{3}}.\label{eq:phi**}
\end{equation}
Here $\text{v}$ is the excluded volume of a monomers of a precursor
polymer in the solution. Chain conformations are ideal for $\phi>\phi^{**}$
and our discussion of the preceding sections applies for these high
volume fractions. Note that the a-thermal limit refers to $\text{v}\approx b^{3}$.

Real polymers occupy space and adjacent monomers of the same polymer
block some of the possible positions for reaction with other reactive
groups. This is taken into account by considering an effective concentration
of reaction partners with the reaction volume, which we denote by
$c_{\text{R}}$. In preceding sections, we have used an internal $c_{\text{i}}$
due to a single reactive group from the same star and $c_{\text{e}}$
for the concentration of external reactive groups on surrounding stars.
For the internal and contributions, one needs to consider the degeneracy
of pairs of open reactive groups for a star with $j$ bonds closed,
equation (\ref{eq:degeneracy}), while only a portion of $1-p$ of
the external groups is available at conversion $p$. This gives for
arbitrary $f$ that 
\begin{equation}
c_{\text{R}}=c_{\text{i}}\sum_{j=0}^{f-2}\text{w}_{\text{j}}D\left(f,j\right)+c_{\text{e}}\left(1-p\right)\label{eq:A_R}
\end{equation}
in the absence of excluded volume. Note that in the limit of high
conversion, $p\rightarrow1$, we expect $\sum_{j=0}^{f-2}\text{w}_{\text{j}}D\left(f,j\right)\rightarrow\text{w}_{\text{f-2}}$.

In the a-thermal limit, we expect in dilute solution, $\phi<\phi^{*}$,
that the contact probability between a single pair of reactive groups
of a star settles around
\begin{equation}
c_{\text{i}}\approx\frac{1}{b^{3}\left(2N/f\right)^{\nu(3+\theta_{0})}}\label{eq:ci-1}
\end{equation}
with contact exponent $\theta_{0}\approx0.273$ for space dimension
$d=3$ \citep{DesCloizeaux1980} and a chain size $b\left(2N/f\right)^{\nu}$
between the reactive groups with $\nu\approx0.587597$ \citep{Clisby2010}
in the a-thermal limit. The above expression uses the end contact
statistics of a long linear chain of $2N/f$ segments (neglecting
corrections due to the star architecture). We consider this as a reasonable
approximation for the limit of low $f$ and large $N$.

For $\phi>\phi^{*}$, we use the scaling approximation of Friedman
and O'Shaugnessy \citep{Friedman1994} that is based upon their renormalization
group calculations \citep{Friedman1988,Friedman1989}. Here, it is
assumed that self-avoiding walk statistics are preserved up to a length
corresponding to the size of a correlation volume and thus, apply
for strands containing roughly
\begin{equation}
g\approx\left(\frac{b^{3}}{\text{v}}\right)^{3(2\nu-1)/(3\nu-1)}\phi^{-1/(3\nu-1)}\label{eq:g}
\end{equation}
segments apart from the end contact. Excluded volume $\text{v}$ is
screened on larger distances than the size $\xi$ of the correlation
volume. Since the size $R$ of a linear strand of $N$ Kuhn segments
scales as $R\propto\xi\left(N/g\right)^{1/2}$ with $\xi\propto bg^{\nu}$
in semi-dilute solutions \citep{Rubinstein2003}, we arrive at
\begin{equation}
c_{\text{i}}\approx\left(\frac{g}{Nb^{2}}\right)^{3/2}g^{-\nu(3+\theta_{0})}\label{eq:ci-2}
\end{equation}
\[
\approx\left[\left(\frac{\text{v}}{b^{3}}\right)^{3(2\nu-1)}\phi\right]^{\left[\nu(3+\theta_{0})-3/2\right]/(3\nu-1)}\frac{1}{b^{3}N^{3/2}}\propto\phi^{0.55}
\]
for semi-dilute good solvent, $\phi^{**}>\phi>\phi^{*}$. For the
dilute case, $\phi<\phi^{*}$, we obtain with $g=N$ that
\begin{equation}
c_{\text{i}}\approx b^{-3}N^{-\nu(3+\theta_{0})}\propto N^{-1.92}.\label{eq:ci-3}
\end{equation}

Similar arguments as above apply for contacts with other chains. Thus,
in the semi-dilute regime, $\phi>\phi^{*}$, we obtain
\begin{equation}
c_{\text{e}}\approx\frac{f\phi}{b^{3}Ng^{\nu\theta_{0}}}\label{eq:ce-1}
\end{equation}
\[
\approx\frac{f}{b^{3}N}\left(\frac{\text{v}}{b^{3}}\right)^{3\nu\theta_{0}(2\nu-1)/(3\nu-1)}\phi^{1+\nu\theta_{0}/(3\nu-1)}\propto\phi^{1.21},
\]
while $g=N$ in the dilute regime, $\phi<\phi^{*}$, leads to 
\begin{equation}
c_{\text{e}}\approx\frac{f\phi}{b^{3}N^{1+\nu\theta_{0}}}.\label{eq:ce}
\end{equation}

The above estimates for $c_{\text{i}}$ and $c_{\text{e}}$ replace
the corresponding expressions in the preceding sections and thus,
allow to compute $\omega$ and $p_{\text{lin}}$ and molecular weight
distributions in good and a-thermal solvents. The general trend that
we obtain is the following: when improving solvent quality, the chains
start to swell in size while increasing their mutual repulsion. This
reduces the contacts between chain ends and the return probability
for cyclization. Thus, conversion and cyclization are both being reduced
with increasing solvent quality, however, at different rates where
the decresae is stronger for the contacts with other chains.

\section{Discussion\label{sec:Discussion}}

We start our discussion by a comparison with preceding work. The approach
of Ercolani et al. \citep{Ercolani1993} is equivalent to our approach
as can be seen from equation (19) of Ref. \citep{Ercolani1993}. In
this equation, the initial monomer concentration is split into three
contributions, one due to strained rings, a second one due to unstrained
rings, and a third one due to linear chains. In our work, we consider
only unstrained rings, since we consider LGS as basic units. In this
case, the first term is dropped. Let us divide out initial monomer
concentration of equation (19) of Ref. \citep{Ercolani1993}. Then
the two remaining terms constitute simply the weight fraction of rings
and chains respectively. Note that $x$ in Ref. \citep{Ercolani1993}
refers to $p_{\text{lin}}$ in our work. The resulting linear chain
contribution is equivalent to our equation (\ref{eq:K}) when considering
that initial monomer concentration is $c_{\text{e}}/2$ and that the
ideal law of mass action, equation (\ref{eq:K}) applies only to the
linear chain fraction - a fact that we have used in section \emph{Linear
polymerization only smallest loop. }Alternatively, one can check that
the weight fraction and distribution of cyclic species in Ref. \citep{Ercolani1993}
and our work are identical. Since both Ref. \citep{Ercolani1993}
and our work solve the equations by a self-consistent recoursion,
all numerical results must be equivalent.

In contrast to Ref. \citep{Ercolani1993}, we do not insert known
results of the most probable distribution of the linear species into
our equations, instead, we have used a set of balance equations that
offers two advantages regarding possible extensions of this mean field
approach. First, by setting up balance equations as a function of
the number of reacted groups, we have paved the way for a direct implementation
of a first shell substitution effect where a reacted (or non-reacted)
group of the same precursor macromonomer implies a different reactivity
of the remaining reactive groups: equation (\ref{eq:pi}) refers to
the simultaneous reaction of both groups, equation (\ref{eq:w1})
deals with the first reactions of a strand, while equation (\ref{eq:w2-1})
and equation (\ref{eq:balances}) for $k>1$ describe the second reaction
of a strand. In case of a first shell substitution effect, only the
modified reaction rates need to be converted into the corresponding
set of modified equilibrium constants for each type of reaction. The
routine for the numerical solution provides $\omega$ and $p_{\text{lin}}$.
However, care must be taken when constructing distributions and averages,
since the frequencies of the smallest molecules are shifted with respect
to a most probable distribution or the FS ring size distribution.
These frequencies are available through $w_{0}$ and equation (\ref{eq:pi})
while the distributions can be constructed after an appropriate renormalization
of $p_{\text{lin}}$ for the remaining molecules with $k\ge2$.

The second advantage is that, we have written down the basic equations
for macromonomers containing an arbitrary number of $f$ reactive
groups. This allows the treatment of reversible networks by a simple
extension of the present work. For such an extension, the smallest
cyclic species made of only a small number of macromonomer units are
of highest interest, since these lead to the largest reduction of
the phantom modulus from the loop free reference \citep{Zhong2016,Lang2018,Lang2019b}.
For irreversible systems, a similar treatment in preceding work \citep{Lang2012a}
can be adapted by replacing the corresponding rate equations by a
simple addition of the corresponding backwards reaction. This extension
of our work is currently in progress.

Another advantage of our approach is that it can be extended to chain
conformations different from a random walk beyond a generalization
of only intramolecular contacts. Deviations from Gaussian conformations
regarding ring closure have been discussed in depth recently \citep{DiStefano2019}.
These were also investigated in several simulation studies \citep{Milchev1994,Chen2004,Lee2018}
with focus on the effect of a different chain stiffness on ring formation
to allow for more precise predictions for a particular polymer. The
general trend is that increasing chain stiffness reduces the amount
of small rings that can be formed, lowering the critical concentration.
These effects are readily incorporated into our treatment similar
to the examples mentioned above: equation (\ref{eq:pi}) and (\ref{eq:lk})
must be multiplied by a $k$-dependent factor that describes the ratio
between the true concentration of the other end of the $k$-mer within
the reaction volume $V_{\text{R}}$ as compared to the $c_{\text{i}}k^{-5/2}$
term of the flexible reference case. Summation over all resulting
ring weight fractions replaces then equation (\ref{eq:wO}).

It is worthwhile to mention that conversion $p$ must not be a monotonic
function of concentration in the cross-over from the ring dominated
to the chain dominated regime \citep{Hagy2007}, which is in agreement
with estimates based upon the Ercolani's work. Our numerical results
fully support this observation, but on top of treating intramolecular
reactions properly, our approach also contains a proper discussion
of the effect of contacts with other chains.

This can be shown by comparing the scaling of the average molecular
weight of the linear chains as a function of $\phi$ in a-thermal
solvent. In the semi-dilute regime, $\phi^{**}>\phi>\phi^{*}$, the
contacts with the reactive groups on other chains are given by equation
(\ref{eq:ce-1}). For sufficiently large volume fractions beyond $\phi^{*}$,
the weight fraction of chains dominates and we approximate $c_{\text{R}}\approx c_{\text{e}}\propto\phi^{1+\nu\theta_{0}/(3\nu-1)}$,
which we insert into equation (\ref{eq:Nn1}). In the limit of large
$K$, this yields 
\begin{equation}
N_{\text{n}}\propto\phi^{1/2+\nu\theta_{0}/(6\nu-2)}\propto\phi^{0.6},\label{eq:Nn good}
\end{equation}
which has been predicted previously for a dominating chain fraction
of worm-like micelles \citep{Cates1990} and which is supported by
the results of Refs. \citep{Wittmer1998,Wittmer2000}. Since this
translates into a concentration dependence of $p$ through $N_{\text{n}}=\left(1-p\right)^{-1}$,
the ring species develops a similarly modified weight distribution
and thus, a slightly modified weight fraction. Since $\phi>\phi^{*}$
in this limit, all precursor macromonomers are above overlap and thus,
the return probability for loop formation is $\propto k^{-3/2}$.
Therefore, the power law part of the weight distributions of rings
is not a function of $\phi$ in the semi-dilute regime. This was also
demonstrated by the simulation data in Ref. \citep{Wittmer2000}.

In the dilute regime $\text{\ensuremath{\phi<\phi^{*}}}$, the weight
fraction of rings dominates. Recall that the weight fraction of rings
resemble the return probability of the corresponding walks. Therefore,
equation (\ref{eq:ci-3}) implies 
\begin{equation}
\omega_{\text{k}}\propto k^{-\nu(3+\theta_{0})}\propto k^{-1.92},\label{eq:omega dilute}
\end{equation}
which leads to $l_{\text{k}}\propto k^{-\nu(3+\theta_{0})-1}\propto k^{-2.92}$.
This latter dependence was derived previously in Ref. \citep{Wittmer2000}
based upon the potts model and good agreement with simulation data
was obtained.

Regarding the linear chains, we recall that the average degree of
polymerization approaches quickly $N_{\text{n}}\approx1$ for $\text{\ensuremath{\phi<\phi^{*}}}$,
as obvious from the quick decay of $p_{\text{lin}}$ for decreasing
$\phi$ in Figure \ref{fig:-as-a}. This stems from the fact that
intramolecular reactions are possible within the smallest unit in
our model system. An analysis of the scaling of $N_{\text{n}}$ in
this limit is not much meaningful. Related work \citep{Wittmer1998,Wittmer2000}
discusses the opposite limit where a significant number of bonds is
necessary to establish the smallest cyclic molecule. The resulting
scaling of the average degree of polymerization
\begin{equation}
N_{\text{n}}\propto\phi^{1/(2+\nu\theta_{0})}\propto\phi^{0.46},\label{eq:Nn dilute}
\end{equation}
indicates the impact of correlations in the surrounding solution that
survive even below the overlap concentration of the molecules.

It is also useful to discuss the full expression for $c_{\text{R}}$

\begin{equation}
c_{\text{R}}\approx\frac{1}{b^{3}Ng^{\nu\theta_{0}}}\left[\frac{\sum_{\text{j}=0}^{f-2}\text{w}_{\text{j}}D\left(f,j\right)}{g^{3\left(\nu-1/2\right)}N^{1/2}}+f\phi\left(1-p\right)\right]\label{eq:c_R}
\end{equation}
that we have written down for the a-thermal limit only for simplication.
The two contributions in the square brackets are due to intra- and
inter-molecular reactions respectively. The contact exponent $\theta_{0}$
appears only in the coefficient in front of this brackets. Therefore,
$\theta_{0}$ impacts reaction kinetics and the equilibrium extent
of reaction, but $\theta_{0}$ does not cause a shift of the relative
contributions of inter- and intra-molecular reactions to conversion.
In other words, the estimate for $\omega$ in the preceding sections
for a given $p$ and $c_{\text{e}}/c_{\text{i}}$ applies universally
for any system with the same $p$ and $c_{\text{e}}/c_{\text{i}}$
that does not phase separate. Note that equation (\ref{eq:c_R}) is
also in line with cyclization data in previous work on irreversible
systems \citep{Lang2005a,Schwenke2011,Lang2012a}. In these works,
reactions were stopped at a defined extent of reaction, which cancels
out any dependence on $p$ and $\theta_{0}$ so that only the terms
inside the square brackets of equation (\ref{eq:c_R}) play a role.
Therefore, it was possible to correctly describe the scaling of the
weight fractions of pending loops etc., without a consideration of
$\theta_{0}$.

Let us now compare with the theoretical treatment of related problems.
Petschek et al. \citep{Petschek1986} discuss a chemical equilibrium
theory for liquid sulphur, which is one of the best studied model
systems of an equilbrium polymerization, see the introduction of Ref.
\citep{Petschek1986} - also regarding other related problems. Here,
two equilibrium constants, $K_{1}$ and $K_{p}$ are needed to describe
initiation and propagation, respectively, since short cyclic sulphur
rings need to break in order to participate in subsequent polymerization
steps. Note that Petschek et. al. \citep{Petschek1986} also allow
for different equilibrium constants of bonds in rings and linear chains.
However, in the strain free case discussed in our work, both equilibrium
constants must be equal to the propagation constant $K_{p}$. By comparing
the expressions for number distribution of chains of Ref \citep{Petschek1986}
with a most probable distribution, we identify $w_{0}K_{p}=p_{\text{lin}}$,
and thus
\begin{equation}
K_{\text{p}}=\frac{p_{\text{lin}}}{\left(1-\omega\right)\left(1-p_{\text{lin}}\right)^{2}}=\frac{p-\omega}{\left(1-p\right)^{2}}\label{eq:Kp}
\end{equation}
while there must be
\begin{equation}
K_{1}\equiv1.\label{eq:K1}
\end{equation}
Thus, reversible polycondensation is a special case of the liquid
sulphur system at a fixed initiation constant of unity.

Most of the discussion in Ref. \citep{Petschek1986} (and preceding
work of the same authors summarized in the same work) deals with the
limit of $K_{1}\rightarrow0$, where the corresponding phase transition
becomes mathematically sharp. Since $K_{1}\equiv1$ for reversible
condensation polymerization, this limit is not accessible, and we
can enforce a vanishing density of chain ends only through the limit
of $K_{\text{p}}\rightarrow\infty$. Petschek et. al. \citep{Petschek1986}
discuss this limit briefly. When increasing concentration at a constant
$K_{\text{p}}$, the weight fraction of linear chains is essentially
zero until reaching a critical point beyond which the weight fraction
of linear chains grows linearly. This is fully equivalent to the discussion
by Ercolani \citep{Ercolani1993} and in our work.

Recently, several models were discussed to estimate the impact of
concatenation on the weight distributions of linear, cyclic and concatenated
molecules in equilibrium polymers \citep{DiStefano2016,DiStefano2017}.
The results of these works indicate that concatenation may be ignored
for flexible polymers, while semi-flexible rings may form predominantly
catenated states that enhance the observed molecular weights. Therefore,
we expect that our analysis should apply well for flexible chain systems,
since these are not disturbed largely by concatenation. However, some
care needs to be taken when analzing semi-flexlible systems, where
the chain stiffness remains not small enough to prevent cyclization.

\section{Summary}

In our contribution, we have presented a simple recursive approach
to treat reversible condensation polymerization with cyclization.
Based upon a minimum set of balance equations, the law of mass action,
Gaussian chain statistics, and the assumption of independent reactions,
exact analytical solutions were derived for systems without cyclization,
for systems containing only smallest loops, or systems that exclusively
form loops. These exact solutions were used to test a numerical scheme
that allows to solve more complex cases. Also, these solutions were
used to check the implementation of Monte Carlo simulations that we
used to double check all numerical solutions of complex systems.

Exact numerical solutions were computed for the general case of a
homopolymerization of flexible precursor polymers. It was shown that
our treatment is equivalent to earlier work in literature that is
less flexible for possible generalizations. Several possible generalizations
have been discussed on a different level of detail. First shell substitution
effects are typically modeled by different balance or rate equations
depending on the state of a precursor macromonomer. Our approach is
based upon the required set of balance equations, thus, implementation
of this effect requires only adjustment of some constants. A generalization
for good solvent is discussed and it is shown that this generalization
requires only an adjustment of cocentrations with respect to contacts
between chains. The resulting scaling of the average molecular weights
agrees with preceding work in the limit of low and high polymer volume
fractions whereby our work provides the advantage of providing an
exact numerical solution (within the approximations made to treat
the chain contacts) over the full range of polymer volume fractions.
Also, key equations have been written down for macromonomers of arbitrary
functionality $f$ to allow a subsequent application of the approach
to non-linear reactions and reversible networks and gels. Finally,
our approach utilizes explicit distributions of ring sizes, which
are the point of attack for implementing polymer specific detail regarding
precursor macromonomers with a low molecular mass or high stiffness.
Therefore, we expect that our approach will find numerous applications
in future research related to supramolecular polymers.

\section{Acknowledgements}

The authors thank the ZIH Dresden for a generous grant of computation
time and the DFG for funding Project LA2735/5-1. The authors also
thank Frank Böhme, Reinhard Scholz, and Toni Müller for useful comments
on earlier versions of the manuscript.

\section{Supporting Information}

The supporting information (SI): Additional Simulation data, Description
of the Monte-Carlo simulations, Description of the numerical solutions,
Minimum example for the recoursive computations necessary to solve
case 1.

\bibliographystyle{apsrev}
\bibliography{library}

\begin{figure*}
\includegraphics[width=1\textwidth]{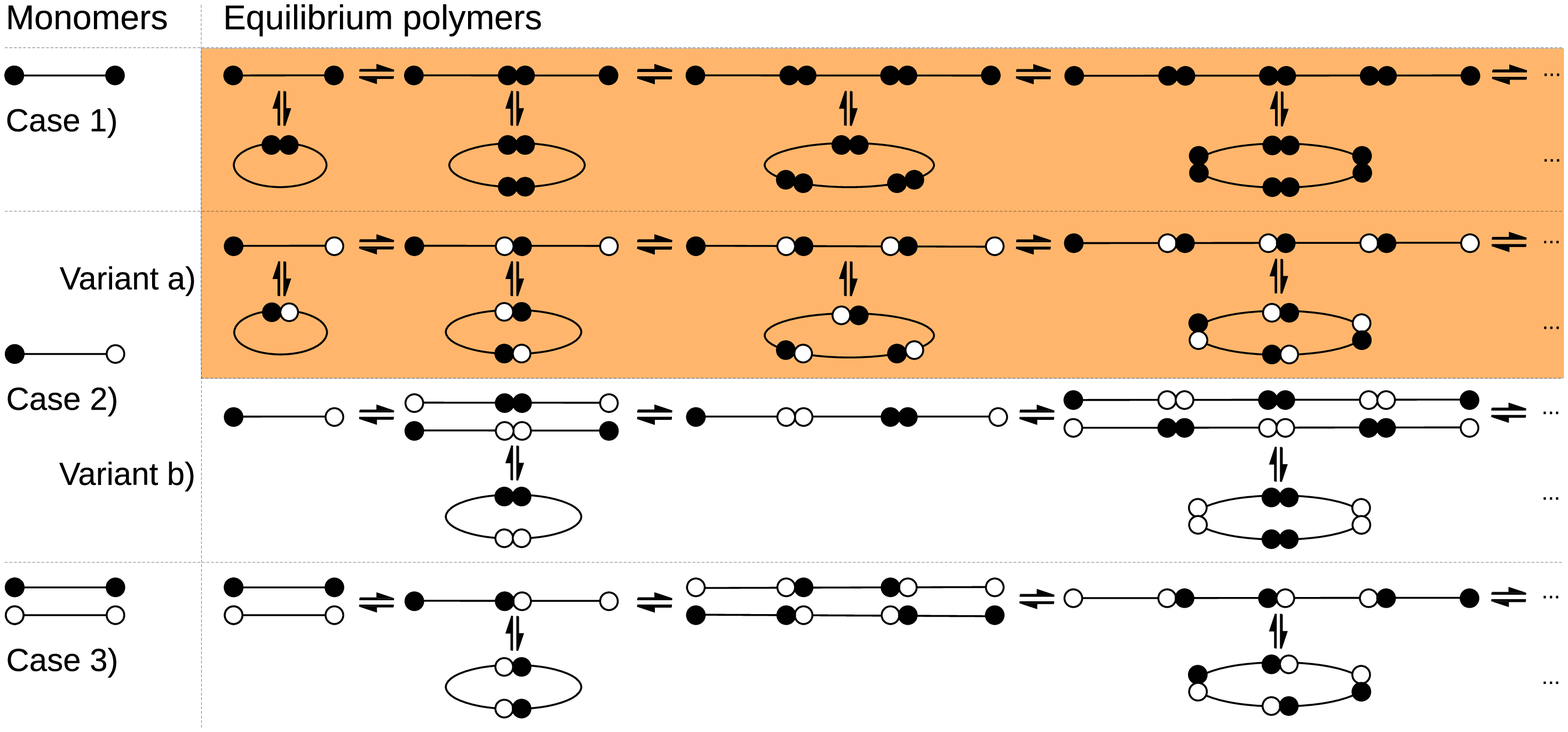}

\caption{Table of contents graphics. Different cases of a stepwise linear condensation
polymerization. Reactive groups of a different type are displayed
by different beads. The different precursor molecules (``monomers'')
on the left assemble into linear and cyclic polymers where the simplest
ones are shown on the right.}
\end{figure*}

\end{document}